\documentstyle[epsf,aps,12pt]{revtex} 
\begin{document} 
\pagestyle{plain} 
\setcounter{page}{1} 
\baselineskip=0.3in 
\begin{titlepage} 
\vspace{.5cm} 
\begin{center} 
{\Large SUSY-QCD Effect on Top-Charm Associated Production 
at Linear Collider     } 

\vspace{.2in} 
Chong Sheng Li $^a$, Xinmin Zhang $^{b}$ and Shou Hua Zhu
$^{c,d}$ \\
$^a$ Department of Physics, Peking University,  
Beijing 100871, P.R. China\\ 
$^b$ Institute of High Energy Physics, P.O. Box 918(4),
 Beijing 100039,
P.R. China\\
$^c$  CCAST (World Lab), P.O. Box 8730, Beijing 100080,
 P.R. China \\
$^d$ Institute of Theoretical Physics, Academia Sinica, P.O. Box
 2735, Beijing 100080, P.R. China
\end{center} 

\begin{footnotesize} 
\begin{center}\begin{minipage}{5in} 
\baselineskip=0.25in 
\begin{center} ABSTRACT\end{center}  
We evaluate the contribution of SUSY-QCD to top-charm associated
production at next generation linear colliders. Our results show
that the production cross section of
the process $e^+e^-\rightarrow t\bar c\mbox{ or }\bar t c$ could be as
large as 
$0.1$ fb, which is larger than the
prediction of the SM by a factor of $10^8$.
\end{minipage}\end{center} 
\end{footnotesize} 
\vfill 
 
 
\end{titlepage} 
 
\newpage 
 
One of the most important physics in top quark sector is to
probe anomalous flavor changing
neutral current (FCNC) couplings.
In the Standard Model (SM), FCNC couplings are forbidden at
the tree level and much suppressed in loops by the GIM
mechanism. Any signals on FCNC couplings in
the
processes of
top
quark decay and productions or indirectly in loops will
indicate the existence of
new physics beyond the SM. Recently in the framework of
effective lagrangian, Han and Hewett\cite{han}
have examined carefully the possibility of exploring
the FCNC couplings $tcZ/ tc \gamma$ in the production vertex
for the reaction $e^+ e^- \rightarrow t {\bar c} + {\bar t}
c$ and concluded that at higher energy colliders with $0.5
-1$ TeV center-of-mass energy, the resulting sensitivity to
FCNC couplings will be
better than the present constraints \cite{zhang}. In this
paper, in the minimal supersymmetric standard model (MSSM)
we study the
process $e^+ e^- \rightarrow t {\bar c}
+ {\bar t}
c$ 
and perform an detail calculation of the contribution from
the FCNC
couplings in the vertex
of gluino-squark-quark 
to the production cross section. We will point out that
at higher energy $e^+ e^-$ colliders 
the cross
section could be as large as 0.1 fb
which is at least eight order of magnitude larger than the
prediction of the SM $\sim 10^{-10} - 10^{-9}$
fb \cite{sm}. 

The MSSM is arguably
the most promising candidate for physics beyond the SM.
Beside many attractive features of supersymmetry in
understanding the mass hierarchy, gauge coupling unification,
the weak scale SUSY models in generally lead to a rich flavor
physics. In fact, SUSY models often have arbitrary flavor
mixings and mass parameters in the squark and slepton sectors
and these mass matrices after diagonalization induce FCNC
couplings at tree level in the vertex of
gluino-squark-quark ${\it etc}$.
Phenomenologically one would have to assume certain
symmetries or dynamical
mechanisms to prevent large FCNC among the first and second
generations. On the other hand the
flavor structure, especially among the second and third
generations in the SUSY sector motivates us to seek for
new physics and any experimental observation on the FCNC
processes beyond the SM would undoubtedly shed light on our
understanding for flavor physics. In this paper we take
model of Ref. \cite{Ellis,Duncan2} where the FCNC couplings
relevant to our calculation is given by:
\begin{eqnarray}
{\cal L_{FC}}=-\sqrt{2}g_sT^aK\overline{\tilde g}P_Lq\tilde{q}_L + h.c.
\end{eqnarray}
In (1), K is the supersymmetric version of the
Kobayashi--Maskawa
matrix, which is explicitly expressed as:
\begin{eqnarray}
K_{ij}=\left(\matrix{1&\varepsilon&\varepsilon^2\cr
 -\varepsilon&1&\varepsilon\cr-\varepsilon^2&-\varepsilon&1\cr}\right)
\end{eqnarray}
where $\epsilon$ parameterizes the strength of flavor mixing
and
is shown to be as large as $1/2$ without
contradicting with the low energy experimental
data \cite{Duncan2}.

In Fig.(1) we give the Feynman diagrams for the
process $e^+(p_1) e^-(p_2)
\rightarrow t(k_1) \bar{c}(k_2)$.
 In
calculations,
we have neglected the scalar u-quark contribution since it is
highly suppressed
by $K_{12}K_{13}$; and we use
the dimensional regularization to control the ultraviolet
divergence. 
We have checked that all divergences cancel out in the final
result with the
summing up of all of the diagrams.
The calculations are carried out in the frame of the center of mass system (CMS)
and Mandelstam variables have been employed:
\begin{eqnarray}
s=(p_1 +p_2)^2 =(k_1 +k_2)^2 \hspace{7mm} t=(p_1 -k_1)^2 \hspace{7mm}
u=(p_1 -k_2)^2.
\end{eqnarray}

After a straightforward calculations, one obtains for the
amplitudes 
\begin{eqnarray}
M&=&{e \over S}\bar{v}(p_1)\gamma_{\mu} u(p_2) \bar{u}(k_1)
V^\mu (tc\gamma) v(k_2) \nonumber \\
&+& {g \over 2 \cos \theta_W (S-M_Z^2)}
\bar{v}(p_1)\gamma_{\mu} (g_V^e-g_A^e \gamma_5) u(p_2) \bar{u}(k_1)
V^\mu (tcZ) v(k_2)
\end{eqnarray}
where,
$g_V^e=1/2-2 \sin^2 \theta_W$, $g_A^e=1/2$, and 
$V^\mu (tc\gamma)$ and $V^\mu (tcZ)$ are the on-shell quarks effective vertices 
given by \footnote{For simplicity, we only give the results in the
limit of $m_c =0$. However in our numerical calculations, we use the
full formulas. }
\begin{eqnarray}
V^\mu(tc\gamma;Z)&=&
f_1^{\gamma;Z} \gamma_\mu P_R+
f_2^{\gamma;Z} \gamma_\mu P_L+
f_3^{\gamma;Z} k_{1\mu} P_R +
f_4^{\gamma;Z} k_{1\mu} P_L 
\nonumber \\
&& + 
f_5^{\gamma;Z} k_{2\mu} P_R +
f_6^{\gamma;Z} k_{2\mu} P_L. 
\end{eqnarray}
The form factors, $f_i^{\gamma;Z}$ are 
\begin{eqnarray}
f_1^\gamma &=&
\sum_{\tilde{q}=
\tilde{c}, \tilde{t}}
{(\pm 1)\epsilon e g_s^2 \cos (\theta_{\tilde{q}}) \sin(\theta_{\tilde{q}}) 
m_{\tilde{g}}
\over 
12 m_t \pi^2 } [
B_0(0, m_{\tilde{g}}^2, m_{\tilde{q}_2}^2)
- B_0(m_t^2, m_{\tilde{g}}^2, m_{\tilde{q}_2}^2)]
+ R.R.
\nonumber \\
f_2^\gamma &=&
\sum_{\tilde{q}=
\tilde{c}, \tilde{t}}
{(\pm 1)\epsilon e g_s^2 \sin^2 (\theta_{\tilde{q}}) \over 
24 m_t^2 \pi^2 } [
 (m_{\tilde{g}}^2-m_{\tilde{q}_2}^2) B_0(0, m_{\tilde{g}}^2, m_{\tilde{q}_2}^2)
- (m_{\tilde{g}}^2-m_{\tilde{q}_2}^2+ m_t^2)
B_0(m_t^2, m_{\tilde{g}}^2, m_{\tilde{q}_2}^2)
\nonumber \\
&&+
4 m_t^2 C_{00} ] 
+ R.R
\nonumber \\
f_3^\gamma &=&
\sum_{\tilde{q}=
\tilde{c}, \tilde{t}}
{(\mp 1)\epsilon e g_s^2 \sin(\theta_{\tilde{q}}) 
\cos (\theta_{\tilde{q}})
m_{\tilde{g}}
 \over 
12  \pi^2 } [ C_0+2 C_1] + R.R.
\nonumber \\
f_4^\gamma &=&
\sum_{\tilde{q}=
\tilde{c}, \tilde{t}}
{(\pm 1)\epsilon e g_s^2 \sin(\theta_{\tilde{q}}) 
\cos (\theta_{\tilde{q}})
m_{\tilde{g}}
 \over 
12  \pi^2 } [ C_0+2 C_2] + R.R.
\nonumber \\
f_5^\gamma &=&
\sum_{\tilde{q}=
\tilde{c}, \tilde{t}}
{(\pm 1)\epsilon e g_s^2 \sin^2(\theta_{\tilde{q}}) 
m_{t}
 \over 
12  \pi^2 } [ C_0+2 C_{11}] + R.R.
\nonumber \\
f_6^\gamma &=&
\sum_{\tilde{q}=
\tilde{c}, \tilde{t}}
{(\mp 1)\epsilon e g_s^2 \sin^2(\theta_{\tilde{q}}) 
m_{t}
 \over 
12  \pi^2 } [ C_0+2 C_{12}] + R.R.
\end{eqnarray}
\begin{eqnarray}
f_1^Z &=&
\sum_{\tilde{q}=
\tilde{c}, \tilde{t}}
{(\mp 1)\epsilon g g_s^2 \sin^2(\theta_w) 
\cos (\theta_{\tilde{q}}) \sin(\theta_{\tilde{q}}) \over 
12 m_t \cos (\theta_w) \pi^2 } [
B_0(0, m_{\tilde{g}}^2, m_{\tilde{q}_2}^2)
- B_0(m_t^2, m_{\tilde{g}}^2, m_{\tilde{q}_2}^2)]
+ R.R.
\nonumber \\
f_2^Z &=&
\sum_{\tilde{q}=
\tilde{c}, \tilde{t}}
{(\mp 1)\epsilon g g_s^2 \sin^2 (\theta_{\tilde{q}}) \over 
96 m_t^2 \cos (\theta_w) \pi^2 } \{
 (-3+4 \sin^2(\theta_w)) [(m_{\tilde{g}}^2-m_{\tilde{q}_2}^2) B_0(0, m_{\tilde{g}}^2, m_{\tilde{q}_2}^2) \nonumber\\
&& - (m_{\tilde{g}}^2-m_{\tilde{q}_2}^2+ m_t^2)
B_0(m_t^2, m_{\tilde{g}}^2, m_{\tilde{q}_2}^2)]
+ 4 m_t^2 (-3 \sin^2 (\theta_{\tilde{q}})+4 \sin^2(\theta_w)) C_{00} \nonumber \\
&& -12 m_t^2 \cos^2 (\theta_{\tilde{q}})
\hat{C}_{00} \}
+ R.R
\nonumber \\
f_3^Z &=&
\sum_{\tilde{q}=
\tilde{c}, \tilde{t}}
{(\pm 1)\epsilon g g_s^2 \sin(\theta_{\tilde{q}}) 
\cos (\theta_{\tilde{q}})
m_{\tilde{g}}
 \over 
48 \cos (\theta_w) \pi^2 } [
(4 \sin^2(\theta_w)-3 \sin^2 (\theta_{\tilde{q}})) ( C_0+2 C_1) \nonumber \\
&& + 3 \sin^2 (\theta_{\tilde{q}}) 
( \hat{C}_0+2 \hat{C}_1) ]  + R.R.
\nonumber \\
f_4^Z &=&
\sum_{\tilde{q}=
\tilde{c}, \tilde{t}}
{(\mp 1)\epsilon g g_s^2 \sin(\theta_{\tilde{q}}) 
\cos (\theta_{\tilde{q}})
m_{\tilde{g}}
 \over 
48  \cos (\theta_w) \pi^2 } [
(4 \sin^2(\theta_w)-3 \sin^2 (\theta_{\tilde{q}})) ( C_0+2 C_2) \nonumber \\
&& + 3 \sin^2 (\theta_{\tilde{q}})
( \hat{C}_0+2 \hat{C}_2) ]  + R.R.
\nonumber \\
f_5^Z &=&
\sum_{\tilde{q}=
\tilde{c}, \tilde{t}}
{(\mp 1)\epsilon g g_s^2 \sin^2(\theta_{\tilde{q}}) 
m_{t}
 \over 
48  \cos (\theta_w) \pi^2 } [
(4 \sin^2(\theta_w)-3 \sin^2 (\theta_{\tilde{q}})) ( C_0+2 C_{11}) \nonumber \\
&& - 3 \cos^2 (\theta_{\tilde{q}})
( \hat{C}_0+2 \hat{C}_{11}) ]  + R.R.
\nonumber \\
f_6^Z &=&
\sum_{\tilde{q}=
\tilde{c}, \tilde{t}}
{(\pm 1)\epsilon g g_s^2 \sin^2(\theta_{\tilde{q}}) 
m_{t}
 \over 
48 \cos (\theta_w) \pi^2 } [
(4 \sin^2(\theta_w)-3 \sin^2 (\theta_{\tilde{q}})) ( C_0+2 C_{12}) \nonumber \\
&& - 3 \cos^2 (\theta_{\tilde{q}})
( \hat{C}_0+2 \hat{C}_{12}) ]  + R.R.
\end{eqnarray}
where $R.R.$ represents the replacement of 
$\theta_{\tilde{q}} \rightarrow \pi/2+ \theta_{\tilde{q}}$
and $m_{\tilde{q}_1} \leftrightarrow m_{\tilde{q}_2}$. 
The variables
of three point
functions $C_{i}$, $C_{ij}$ \cite{denner} and
$\hat{C}_{i}$, $\hat{C}_{ij}$
are 
$(m_t^2, S, 0, m_{\tilde{g}}^2,m_{\tilde{q}_2}^2, m_{\tilde{q}_2}^2)$ 
 and $(m_t^2, S, 0, m_{\tilde{g}}^2,m_{\tilde{q}_2}^2, m_{\tilde{q}_1}^2)$,
 respectively.


 In the MSSM the mass eigenstates of the 
squarks $\tilde{q}_1$ and $\tilde{q}_2$ are related to the weak
eigenstates $\tilde{q}_L$ and
$\tilde{q}_R$ 
by \cite{MSSM}
\begin{eqnarray}
\left(\begin{array}{c}
\tilde{q}_1 \\ \tilde{q}_2\end{array}\right)=
R^{\tilde{q}}\left(\begin{array}{c}
\tilde{q}_L \\ \tilde{q}_R\end{array}\right)\ \ \ \ \mbox{with}\ \ \ \
R^{\tilde{q}}=\left(\begin{array}{cc}
                   \cos\theta_{\tilde{q}} & \sin\theta_{\tilde{q}}\\
                   -\sin\theta_{\tilde{q}} & \cos\theta_{\tilde{q}}
                   \end{array}
                   \right).
\label{eq1}
\end{eqnarray}
For the squarks, the mixing angle $\theta_{\tilde{q}}$ and 
the masses $m_{\tilde{q}_{1,2}}$ can be calculated by 
diagonalizing the following mass
matrices
\begin{eqnarray}
M^2_{\tilde{q}}=\left(\begin{array}{cc}
          M_{LL}^2 & m_q M_{LR}\\
           m_q M_{RL} & M_{RR}^2
           \end{array} \right), \nonumber \\
M_{LL}^2=m_{\tilde{Q}}^2+m_q^2+m_{z}^2\cos 2\beta (I_q^{3L}-e_q\sin^2\theta_w),
\nonumber \\
M_{RR}^2= m_{\tilde{U},\tilde{D}}^2 +m_q^2+m_{z}^2\cos 2\beta e_q\sin^2\theta_w,
 \nonumber \\
M_{LR}= M_{RL}=\left\{ \begin{array}{ll}
                A_t-\mu \cot \beta & (\tilde{q}= \tilde{t})\\
                A_b-\mu \tan \beta & (\tilde{q}= \tilde{b}),
                \end{array}
                \right.
\label{eq2}
\end{eqnarray}
where $ m_{\tilde{Q}}^2$, $ m_{\tilde{U},\tilde{D}}^2$ are
 soft SUSY breaking mass terms of the left- and right-handed
 squark, respectively; $\mu$ is the
coefficient of the $H_1H_2$
term in the superpotential;
 $A_t$ and $A_b$ are the
coefficient of the dimension-three tri-linear 
soft SUSY-breaking terms; $I_q^{3L}, e_q$ are the weak
isospin
and electric charge of the squark $\tilde{q}$.
From Eqs. \ref{eq1} and \ref{eq2},
we have
\begin{eqnarray}
m^2_{\tilde{t}_{1,2}}&=&{1\over 2}\left[ M^2_{LL}+
M^2_{RR}\mp \sqrt{
(M^2_{LL}-M^2_{RR})^2+4 m^2_t M^2_{LR}}\right] \nonumber \\
\tan\theta_{\tilde{t}}&=&{m^2_{\tilde{t}_1}-M^2_{LL} \over m_t M_{LR}}.
\label{eq3}
\end{eqnarray}

 Now we present the numerical results.
For the SM parameters, we take
\begin{eqnarray}
m_Z=91.187 GeV, ~~~~m_W=80.33 GeV&&, ~~~~ m_t=176.0 GeV, ~~~~
m_c=1.4 GeV
\nonumber \\
\alpha=1/128 &&, ~~~~ \alpha_S=0.118
\end{eqnarray}
For the MSSM parameters, we choose $\mu=-100 GeV$ and
$\epsilon^2 =
1/4$.
To simplify the calculation we have taken 
that  $m_{\tilde{U}} = m_{\tilde
{D}}=m_{\tilde{Q}}=A_t=m_S$ (global SUSY).
In Figs. 2-5, we show the cross sections of the process
$e^+e^- \rightarrow t \bar c$ as functions of $m_S$, 
$m_{\tilde{g}}$,
$\sqrt{s}$ and $\tan\beta$. One can see that
the production cross section increases as squarks  and
gluino masses decrease, and it could reach $0.1 $ fb for
favorable parameters. This is an enhancement by a factor of
$10^8$
relative to the SM prediction. 
 Such enhancement could be easily understood as following:
\begin{eqnarray}
\frac{\sigma_{SUSY}}{\sigma_{SM}}
\sim \left( \frac{\alpha_s \Delta m_{\tilde{q}}^2}{\alpha m_b^2} \right)^2,
\end{eqnarray}
where $\Delta m_{\tilde{q}}^2$ represents the possible mass square difference
among squarks. If $\Delta m_{\tilde{q}}^2$ varies from $100^2 -200^2 (GeV)^2$,
$\frac{\sigma_{SUSY}}{\sigma_{SM}}= 10^7 \sim 10^8$. At the same
time, this kind of enhancement could also be observed
in FCNC decay process of top quark 
\cite{decay}.
Due to the
rather clean experimental environment and well-constrained
kinematics, the signal of ${\bar t} c$ or $t {\bar c}$
would be spectacular \cite{han}. We expect the
SUSY-QCD effects studied in this paper be observed at higher
energy $e^+ e^-$ colliders.
    
\section*{Acknowledgments} 
We thank Profs. Tao Han and Chao-Shang Huang for discussions.
This work was supported in part by the
 National Natural Science Foundation
of China,
Doctoral Program Foundation of
Higher Education,
the post doctoral foundation
of China,
 and a grant from the State Commission of Science and
 Technology of China.
 S.H. Zhu also
 gratefully acknowledges the
 support of  K.C. Wong Education Foundation, Hong Kong.


\begin{figure}
\epsfxsize=14 cm
\centerline{\epsffile{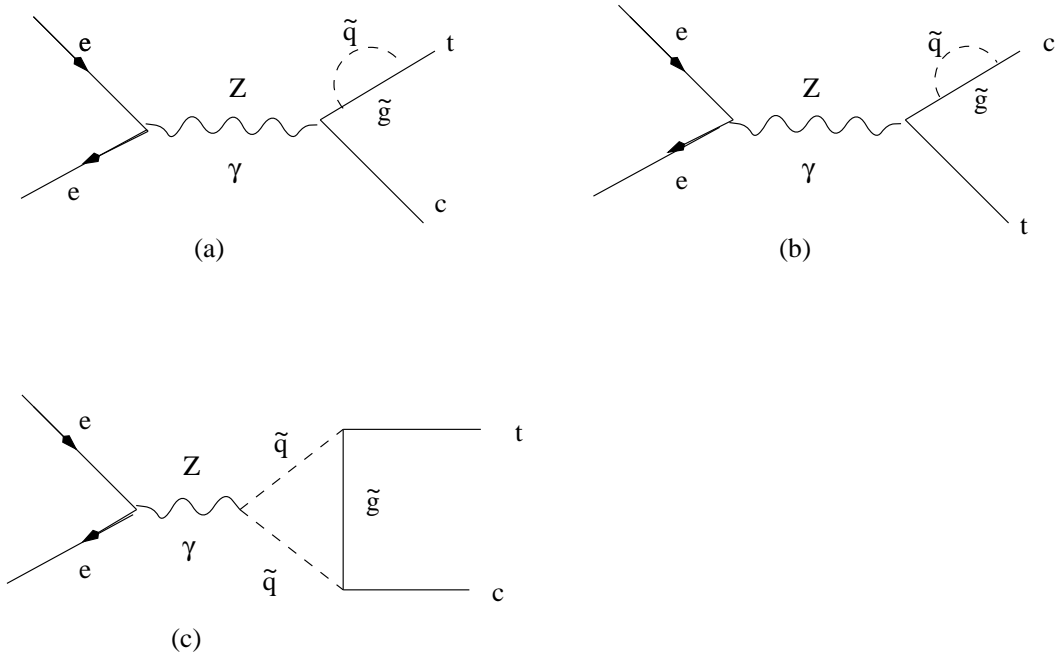}}
\caption[]{
The Feynmann diagrams for the process $e^+ e^- \rightarrow t \bar{c}$.
}
\end{figure}

\begin{figure} 
\epsfxsize=14 cm 
\centerline{\epsffile{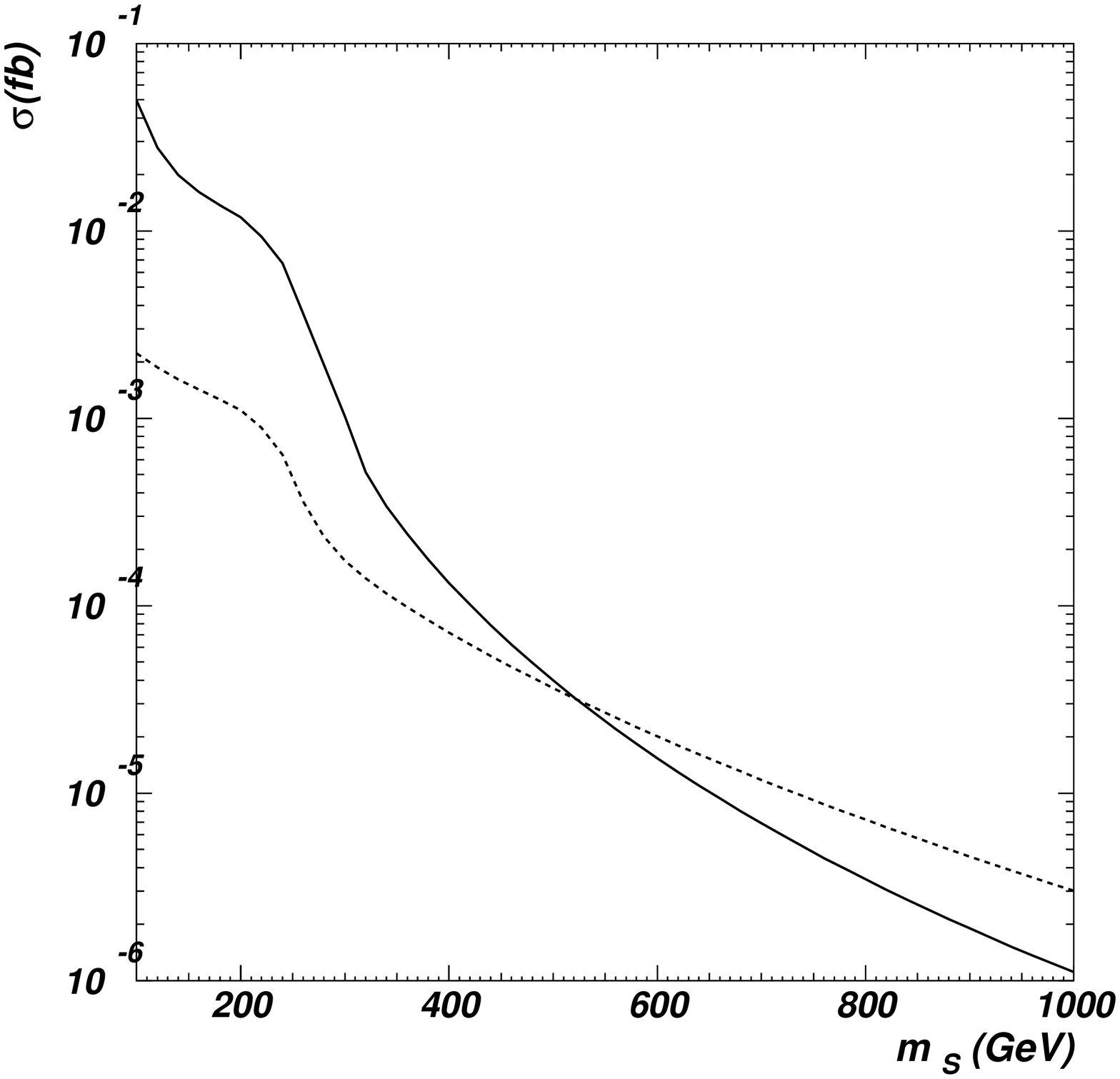}} 
\caption[]{
The cross section for the process $e^+ e^- \rightarrow t \bar{c}$
as 
a function  of  $m_S$,
where $\sqrt{S}= 500 GeV$,  $\tan\beta =2 $,
$\epsilon^2=1/4$ and $\mu =-100 GeV$. The solid  and dashed lines
represent $m_{\tilde{g}}= 100 GeV$ and  $500 GeV$, respectively.
} 
\end{figure} 

\begin{figure} 
\epsfxsize=14 cm 
\centerline{\epsffile{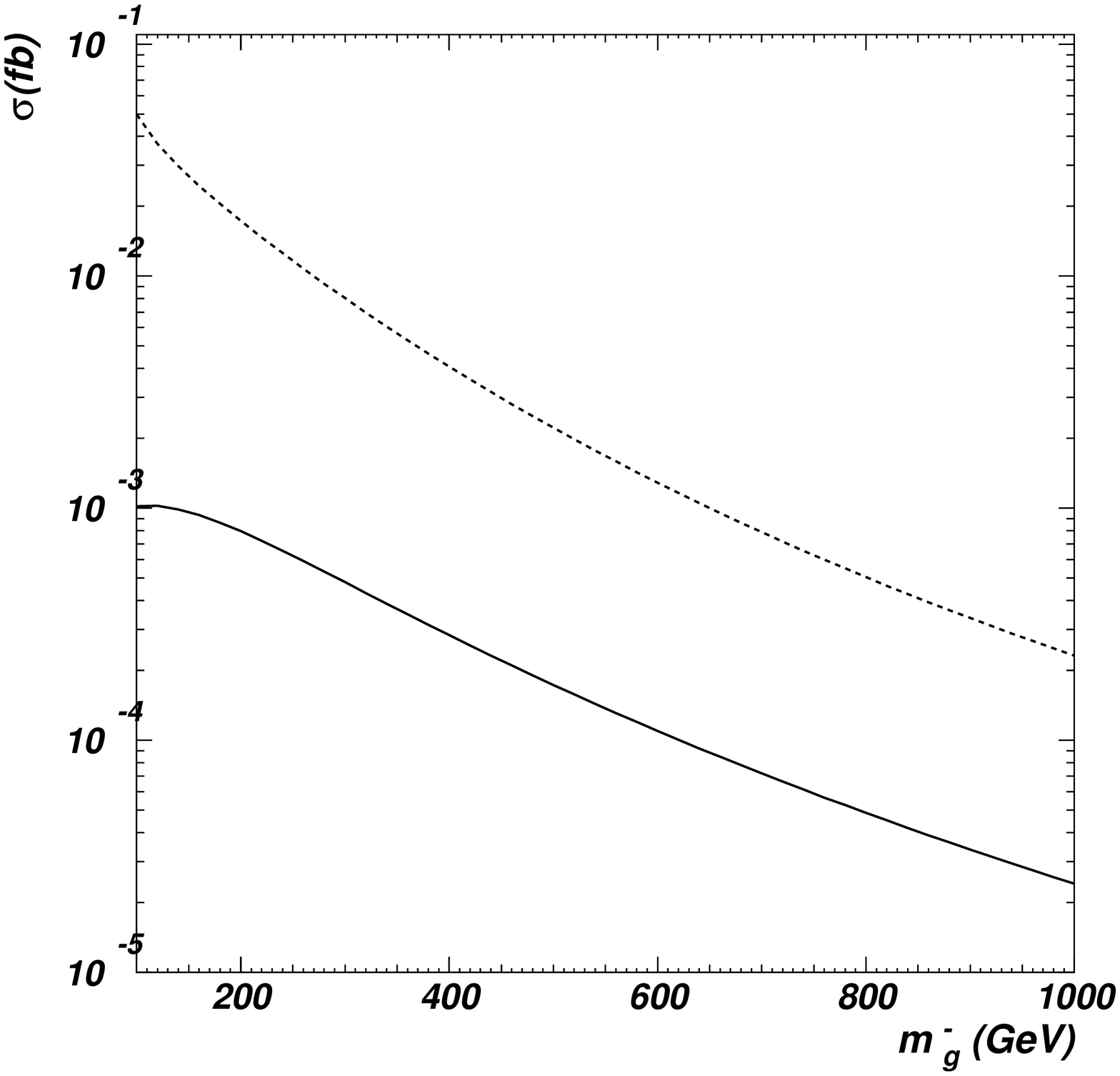}} 
\caption[]{
The cross section for the process $e^+ e^- \rightarrow t \bar{c}$
as 
a function  of  $m_{\tilde{g}}$,
where $\sqrt{S}= 500 GeV$,  $\tan\beta =2 $,
$\epsilon^2=1/4$ and $\mu =-100 GeV$.
The solid  and dashed lines
represent $m_S= 300 GeV$ and  $100 GeV$, respectively.
}
\end{figure} 
 
\begin{figure} 
\epsfxsize=14 cm 
\centerline{\epsffile{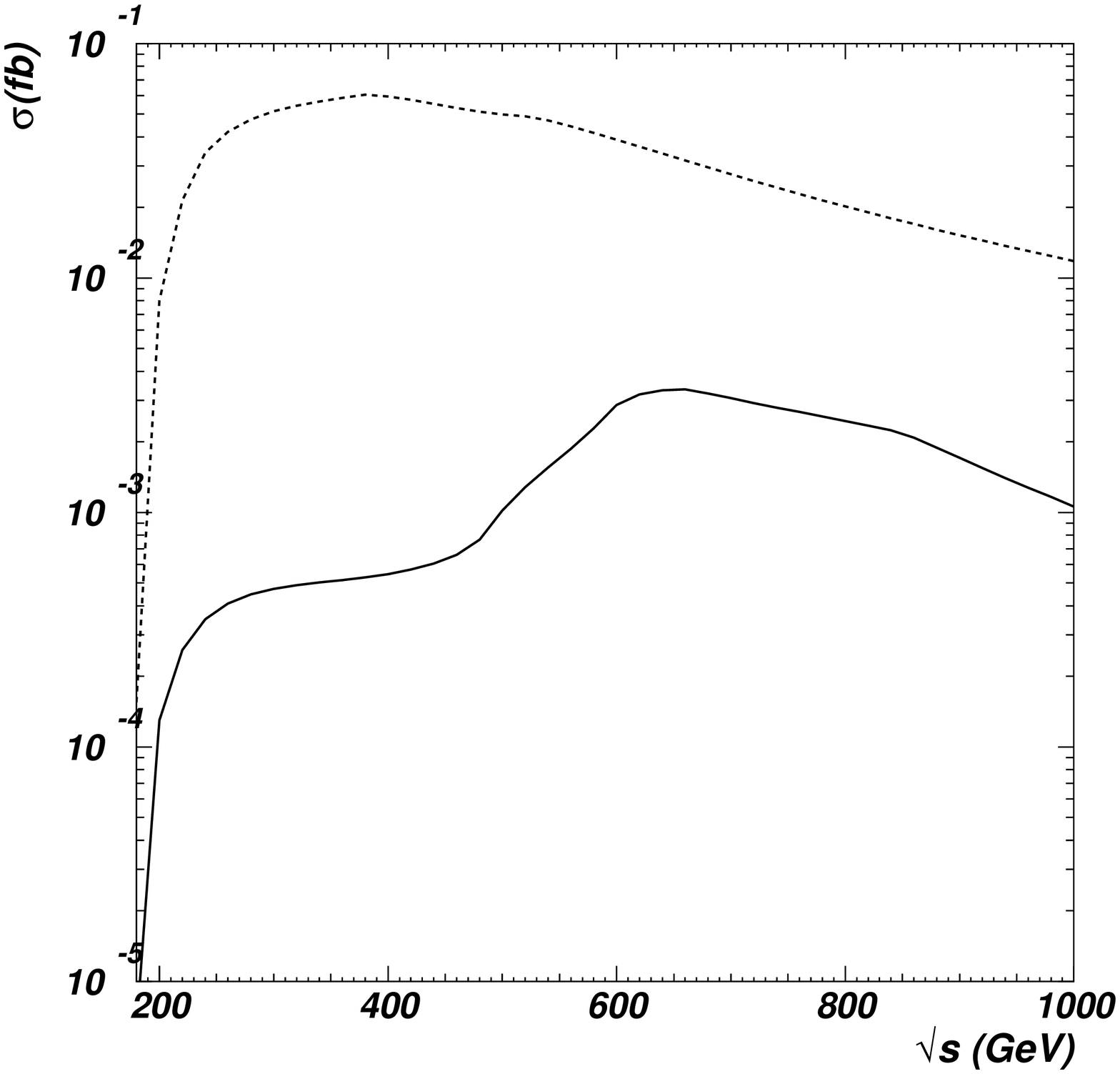}} 
\caption[]{
The cross section for the process $e^+ e^- \rightarrow t \bar{c}$
as 
a function  of  $\sqrt{S}$,
where $m_{\tilde{g}}= 100 GeV$,  $\tan\beta =2 $,
$\epsilon^2=1/4$ and $\mu =-100 GeV$. The solid and dashed lines represent
$m_S= 300 GeV$ and $100 GeV$, respectively.
} 
\end{figure} 

\begin{figure} 
\epsfxsize=14 cm 
\centerline{\epsffile{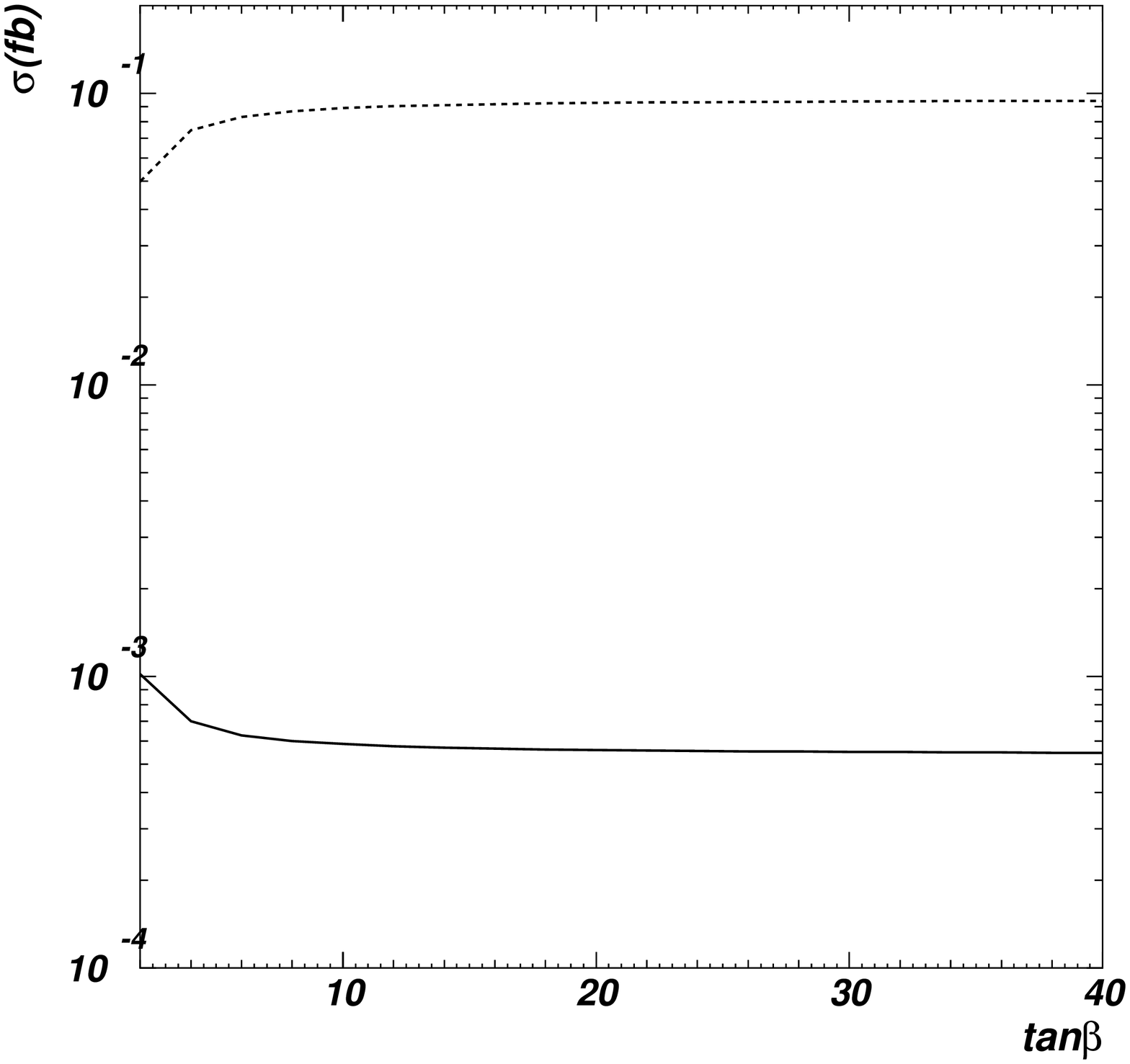}} 
\caption[]{
The cross section for the process $e^+ e^- \rightarrow t \bar{c}$
as 
a function  of  $\tan\beta$,
where $\sqrt{S}= 500 GeV$, $m_{\tilde{g}}= 100 GeV$, 
$\epsilon^2=1/4$ and $\mu =-100 GeV$.
The solid and dashed lines represent
$m_S= 300 GeV$ and $100 GeV$, respectively.
} 
\end{figure} 


\end{document}